\input harvmac
\noblackbox

\font\ticp=cmcsc10
 
\def\Title#1#2{\rightline{#1}\ifx\answ\bigans\nopagenumbers\pageno0\vskip1in
\else\pageno1\vskip.8in\fi \centerline{\titlefont #2}\vskip .5in}

\font\ticp=cmcsc10
\font\ttsmall=cmtt10 at 8pt

\input epsf
\ifx\epsfbox\UnDeFiNeD\message{(NO epsf.tex, FIGURES WILL BE
IGNORED)}
\def\figin#1{\vskip2in}
\else\message{(FIGURES WILL BE INCLUDED)}\def\figin#1{#1}\fi
\def\ifig#1#2#3{\xdef#1{Fig.~\the\figno}
\goodbreak\topinsert\figin{\centerline{#3}}%
\smallskip\centerline{\vbox{\baselineskip12pt
\advance\hsize by -1truein\noindent{\bf Fig.~\the\figno:} #2}}
\bigskip\endinsert\global\advance\figno by1}

%
%

\def\BH{black hole}

\def\SAdS{Schwarzschild-AdS}
\def\AfSf{$AdS_5 \times S^5$}
\def\AfSs{$AdS_4 \times S^7$}
\def\AsSf{$AdS_7 \times S^4$}

\def\[{\left [}
\def\]{\right ]}
\def\({\left (}
\def\){\right )}
\def\g0{\buildrel \circ \over g}

\def\eps{\varepsilon}


\lref\smolin{L. Smolin, hep-th/0003056}
\lref\juan{J. Maldacena, Adv. Theor. Math. Phys. 2 (1998) 231, hep-th/9711200.}
\lref\magoo{O. Aharony, S.S. Gubser, J. Maldacena, H. Ooguri, Y. Oz,
Phys. Rept. 323 (2000) 183, hep-th/9905111.}
\lref\gary{G. Horowitz, Class. Quant. Grav. 17 (2000) 1107, hep-th/9910082.}
\lref\grla{R. Gregory and R. Laflamme, Phys. Rev. Lett. 70 (1993) 2837, 
hep-th/9301052; Nucl. Phys. B428 (1994) 399, 
hep-th/9404071.} 
\lref\hawk{S. Hawking, Phys. Rev. D14 (1976) 2460.}
\lref\wald{R. M. Wald,``General Relativity,''  U. of Chicago Press (1984).}
\lref\ernst{F. J. Ernst, J. Math. Phys. 17 (1976) 54.}

\baselineskip 16pt
\Title{\vbox{\baselineskip12pt
\line{\hfil \tt hep-th/0005288} }}
{\vbox{
{\centerline{Note on Small Black Holes in $AdS_p\times S^q$}}
}}
\centerline{\ticp Gary T. Horowitz and Veronika E. Hubeny\footnote{}{\ttsmall
gary@cosmic.physics.ucsb.edu, veronika@cosmic.physics.ucsb.edu}}
\bigskip
\centerline{\it Physics Department, University of California,
Santa Barbara, CA 93106, USA}
\bigskip
\centerline{\bf Abstract}
It is commonly believed that small black holes in $AdS_5\times S^5$
can be described by the ten dimensional Schwarzschild solution.
This requires that the self-dual five-form (which is nonzero in the
background) does not fall through the horizon and cause the black hole to grow.
We verify that this is indeed the case: There are static solutions to the
five-form field equations in a ten dimensional Schwarzschild spacetime.
Similar results hold for other
backgrounds $AdS_p\times S^q$ of interest in supergravity.
\bigskip

\Date{June, 2000}

%

One of the most important consequences of the AdS/CFT correspondence
\refs{\juan,\magoo}
is the claim that the formation and evaporation of  black holes can be
described by a standard unitary evolution. 
Since this claim is contrary to well known
semiclassical arguments \hawk,
it is worthwhile to carefully examine the
ingredients which go into this conclusion. One such ingredient is the
assumption that a small black hole in \AfSf\ will behave just like a ten
dimensional Schwarzschild black hole. 
Intuitively, this seems reasonable since for a large
AdS radius, the local description should be approximately 
given by the corresponding 
flat ten-dimensional  spacetime physics; 
 in particular, a small black hole should be  
approximately described by the 10-D Schwarzschild solution
(sufficiently near the \BH).

However, the supergravity solution also includes a nonzero five-form. Although
this acts like a cosmological constant in solutions which are products of
two five dimensional spaces, in general it contains dynamical degrees of 
freedom. Given our experience with previous `no-hair' theorems, one might
worry that a small black hole will cause the five-form to fall into
the horizon. Even though the local energy density
in the five-form is small, if this were the case, most small black 
holes would grow by classically absorbing the energy density of the
five-form and not quantum mechanically evaporate.  

We show below that this does not occur. There exist static solutions for
a self-dual five-form in the background of a ten dimensional black hole which
have the correct boundary conditions at infinity to match onto the \AfSf\
solution. It is these boundary conditions which effectively stabilize the
field and invalidate the `no-hair' intuition. The five-form is distorted by
the black hole, but does not cause it to grow. We also  show that similar
results hold for four-forms 
and seven-forms in the background of an 11-D Schwarzschild solution
with the right boundary conditions to match onto \AfSs\ and \AsSf.

We start by noting that for pure \AfSf, the solution 
(in global coordinates) is given by the metric
\eqn\AdS{
ds^2  \, = -\({\rho^2 \over R^2} + 1\) \, dt^2 + 
{d\rho^2 \over {\rho^2 \over R^2} + 1}
+ \rho^2 \, d\Omega_3^2
+ d\chi^2 + R^2 \, \sin^2 {\chi \over R}  \, d\Omega_4^2
}
where $R$ is the radius of curvature, 
and the five-form field strength $\tilde F$ is the sum of the volume form on 
$AdS_5$ and on $S^5$, normalized so that $\int_{S^5} \tilde F =N$. To simplify
the formulas below, we will work with the rescaled five-form
$F\equiv (\pi^3 R^5/N)
\tilde F$, so that $F$ is just the sum of the volume forms\foot{
We use $d\Omega_n$ to denote the volume $n$-form on unit $S^n$ and 
$d\Omega_n^2$ to denote the metric on  $S^n$.}
\eqn\F{
F = - \rho^3 \, dt \wedge d\rho \wedge  d\Omega_3
+ R^4 \, \sin^4 {\chi \over R} \, d\chi \wedge  d\Omega_4
}

How does this solution change
in the presence of a black hole?
For a \BH\ with radius larger than $R$,
we already know what this modification is:
The metric on $AdS_5$
is replaced with the five-dimensional \SAdS\ solution and the
metric on the $S^5$ is unchanged 
\eqn\SadS{
ds^2  \, = -\({\rho^2 \over R^2} + 1 - {\rho_0^2 \over \rho^2}\) \, dt^2 + 
{d\rho^2 \over {\rho^2 \over R^2} + 1 - {\rho_0^2 \over \rho^2}}
+ \rho^2 \, d\Omega_3^2
+ d\chi^2 + R^2 \, \sin^2 {\chi \over R}  \, d\Omega_4^2
}
Since this change in the metric does not effect the volume form on AdS, 
the five-form field strength $F$ remains the same.
In particular, the self-duality condition is satisfied because 
only the combination $dt \wedge d\rho$ is present in this condition,
so that the mass-dependence cancels out, 
and the ``Bianchi identity'' $dF=0$ is independent of the metric.
(It is clear that $F$ remains smooth even at the horizon since the 
volume form on \SAdS\ is smooth there.)

For a small \BH, the picture becomes much less clear.
The \BH\ is localized on the $S^5$ as well as in the $AdS_5$ \grla,
so that the metric no longer factorizes.
Hence we cannot just look for a lower dimensional solution with an
effective cosmological constant.  
Finding the appropriate exact solution to the full 10-D Einstein five-form
field equations
seems intractable. Since the curvature near the horizon of a small black hole
should be much larger than the field strength $F$, to a good approximation
one can ignore the backreaction
and treat the five-form as a test field on a fixed 
background spacetime. In this approximation, the metric satisfies the
vacuum equations, and the unique static, spherically symmetric
black hole solution is
the ten-dimensional Schwarzschild metric.
However, this approximation is consistent only if there exists a static solution
for a test self-dual five-form in this background, with the right boundary 
conditions. These boundary conditions can be understood as follows.

Very far away from the \BH, both the metric and the five-form should approach
the forms given respectively by  eqs.\ \AdS\ and \F. Since the black hole
is much smaller than $R$, these forms are valid even into the approximately flat
region of small $\rho$ and $\chi$. We can identify this approximately flat
region with the asymptotic region far from the Schwarzschild black hole.
This then sets our boundary conditions ``at infinity''.
To be more explicit,
we first write the 10-D Schwarzschild solution in convenient
coordinates in which the boundary conditions are easily posed 
while the required symmetries are still manifest.
In particular, we want to use the 10-D radial coordinate 
(fixed by the area of $S^8$), but to split 
$S^8$ into $S^3$ and $S^4$, corresponding to the rotational SO(4)
symmetry of $AdS_5$ and the remaining (unbroken)
 SO(5) rotational symmetry on $S^5$.
This is achieved by using the coordinate transformation 
$$\rho = r \sin \theta$$
\vskip -0.3 in
\eqn\rhochi{\chi = r \cos \theta}
In these coordinates, the flat spacetime metric obtained from \AdS\ in the
limit $\rho,\chi \ll R$ takes the form
\eqn\flatr{
ds^2  \, = - dt^2 + dr^2
+ r^2 \, \( d\theta^2 + \sin^2 \theta \, d\Omega_{3}^2
+ \cos^2 \theta \,  d\Omega_{4}^2 \)
}
(The angular term in the parentheses is equivalent to 
$d\Omega_8^2$.)
Similarly, the five-form field strength obtained from \F\ and \rhochi\
in the limit
$\rho,\chi \ll R$ takes the form 
$$F = - r^{3} \, \sin^{4} \theta \,
dt \wedge dr \wedge  d\Omega_{3}
-  r^{4} \, \sin^{3} \theta \, \cos \theta \,
 dt \wedge d\theta \wedge  d\Omega_{3}$$
\vskip -0.3 in
\eqn\Fr{
+ r^{4} \,  \cos^{5} \theta \, dr \wedge  d\Omega_{4}
 - r^{5} \, \sin \theta \, \cos^{4} \theta \,
 d\theta \wedge  d\Omega_{4}
}
One can easily recheck that $F$ is still closed and self-dual.

In these coordinates, the 10-D
Schwarzschild metric is given by:
\eqn\schwr{
ds^2  \, = - f(r)  \,  dt^2 + f^{-1}(r)  \,  dr^2
+ r^2 \, \( d\theta^2 + \sin^2 \theta \, d\Omega_{3}^2
+ \cos^2 \theta \,  d\Omega_{4}^2 \)
}
with  $f(r) \equiv 1 - \( {r_+ \over r} \)^{7}$.
A general ansatz for the field strength with the 
required symmetries 
(namely $F$ being  static and spherically symmetric on $S^3$ and $S^4$) 
can be obtained by taking each of the four terms in \Fr\ and multiplying
by arbitrary functions of $r$ and $\theta$:
$$F = - g_1(r,\theta) \, r^{3} \, \sin^{4} \theta \,
dt \wedge dr \wedge  d\Omega_{3}
- g_3(r,\theta) \,  r^{4} \, \sin^{3} \theta \, \cos \theta \,
 dt \wedge d\theta \wedge  d\Omega_{3}$$
\vskip -0.3 in
\eqn\Fgr{
+ g_2(r,\theta) \, r^{4} \,  \cos^{5} \theta \, dr \wedge  d\Omega_{4}
 - g_4(r,\theta) \, r^{5} \, \sin \theta \, \cos^{4} \theta \,
 d\theta \wedge  d\Omega_{4}
}
Our boundary conditions require that
$g_i(r,\theta) \to 1$ as $r \to \infty$.\foot{
In principle, terms of the form
$\gamma_1(r,\theta) \,  dt \wedge d\Omega_{4}$ and
$\gamma_2(r,\theta) \,  dr  \wedge d\theta \wedge  d\Omega_{3}$
would also be consistent with all the symmetries, 
and would satisfy
the boundary conditions provided
$\gamma_i(r,\theta) \to 0$ as $r \to \infty$.
However, since F is closed,
$\partial_r \gamma_1 = \partial_{\theta} \gamma_1 = 0$,
which, along with the boundary condition $\gamma_1 \to 0$, requires 
that $\gamma_1(r,\theta) \equiv 0$.  
Self duality of $F$ then forces $\gamma_2(r,\theta) \equiv 0$.  
Hence these terms will not arise, and the most general form of
$F$ will indeed be given by \Fgr.}
To determine the field strength \Fgr\ explicitly, we now
impose the physical conditions that $F$ is closed and self-dual 
(with respect to the black hole metric \schwr).

We first eliminate two of the four arbitrary functions $g_i$ 
appearing in \Fgr\ by imposing self-duality, $F = \ast F$.
The volume form associated with \schwr\ is simply given by
\eqn\epsr{
\eps_{(10)} =  r^8 \, \sin^3\theta \, \cos^4\theta \, 
 dt \wedge dr \wedge d\theta  \wedge d\Omega_3 \wedge  d\Omega_4
}
Correspondingly, the dual of $F$ is 
$$\ast F = - g_1(r,\theta) \, r^{5} \, \sin \theta \, \cos^{4} \theta \,
 d\theta \wedge  d\Omega_{4}
+ g_3(r,\theta) \, {1 \over f(r)} \, r^{4} \,  \cos^{5} \theta \, 
dr \wedge  d\Omega_{4}$$
\vskip -0.4 in
\eqn\dFgr{
- g_2(r,\theta) \, f(r)  \,  r^{4} \, \sin^{3} \theta \, \cos \theta \,
 dt \wedge d\theta \wedge  d\Omega_{3}
 - g_4(r,\theta) \, r^{3} \, \sin^{4} \theta \,
dt \wedge dr \wedge  d\Omega_{3}
}
Self-duality then requires $g_4 = g_1$ and $g_3 = f g_2$, so
that \Fgr\ becomes
$$ F =  g_1(r,\theta) \, \[ - r^{3} \, \sin^{4} \theta \,
dt \wedge dr \wedge  d\Omega_{3}
- r^{5} \, \sin \theta \, \cos^{4} \theta \,
 d\theta \wedge  d\Omega_{4} \] $$
\vskip -0.3 in
\eqn\Fgdr{
+ g_2(r,\theta) \, \[ -  r^{4}  \, f(r) \, \sin^{3} \theta \, \cos \theta \,
 dt \wedge d\theta \wedge  d\Omega_{3}
+ r^{4} \,  \cos^{5} \theta \, dr \wedge  d\Omega_{4} \]
}
The condition that $F$ is nonsingular at the horizon requires that
the arbitrary functions $g_1(r,\theta)$ and $g_2(r,\theta)$ are smooth
at $r_+$.  This can be easily seen by
rewriting \Fgdr\ in the ingoing Eddington coordinates, 
$(v,r,\theta,\Omega_3,\Omega_4)$,
which are regular at the horizon.  
Since $v \equiv t + r_{\ast}$, where $ r_{\ast}$ is defined by
$d r_{\ast} \equiv {dr \over f(r)}$, we can simply rewrite 
$dt=dv-{dr \over f(r)}$.
The field strength is then expressed as 
$$ F =  g_1(r,\theta) \, \[ - r^{3} \, \sin^{4} \theta \,
dv \wedge dr \wedge  d\Omega_{3}
- r^{5} \, \sin \theta \, \cos^{4} \theta \,
 d\theta \wedge  d\Omega_{4} \] $$
\vskip -0.4 in
$$+ g_2(r,\theta) \, [ -  r^{4}  \, f(r) \, \sin^{3} \theta \, \cos \theta \,
 dv \wedge d\theta \wedge  d\Omega_{3}
+   r^{4} \, \sin^{3} \theta \, \cos \theta \,
 dr \wedge d\theta \wedge  d\Omega_{3}$$
\vskip -0.3 in
\eqn\Fgdrv{
+ r^{4} \,  \cos^{5} \theta \, dr \wedge  d\Omega_{4} ]
}
and we see that all the terms are smooth at $r_+$ 
if  $g_i$ are smooth at $r_+$.

We now require that $F$ is closed, $dF=0$.
Since $dF$ has two nontrivial components,  proportional to
$dt \wedge dr \wedge d\theta \wedge  d\Omega_{3}$ and to
$dr \wedge d\theta \wedge  d\Omega_{4}$, we obtain two independent equations
by setting each component to $0$:
\eqn\closedA{
r^3 \, \partial_{\theta} (g_1 \, \sin^4 \theta) 
- \partial_r (r^4 \, f \, g_2) \, \sin^3\theta \, \cos \theta = 0
}
\eqn\closedB{
\partial_r (r^5 \, g_1) \, \sin \theta \, \cos^4 \theta 
+ r^4 \, \partial_{\theta} (g_2 \, \cos^5 \theta) = 0
}
We can simplify these partial differential equations further by 
separation of variables.
By writing $g_i(r,\theta) \equiv g_i(r) \, {\tilde g}_i(\theta)$,
the radial and angular parts decouple.
By direct substitution, \closedA\ becomes
\eqn\sepA{
{ {\tilde g}'_1(\theta) \, \tan \theta + 
4 {\tilde g}_1(\theta) \over {\tilde g}_2(\theta)}
= k =
{r  \, f(r)  \, g'_2(r)  + 4 f(r)  \, g_2(r) + r  \, f'(r)  \, g_2(r) \over  g_1(r)}
}
whereas  \closedB\ yields
\eqn\sepB{
{ {- \tilde g}'_2(\theta) \, \cot \theta + 
5 {\tilde g}_2(\theta) \over {\tilde g}_1(\theta)}
= l =
{r  \, g'_1(r)  + 5 g_1(r)  \over  g_2(r)}
}
where $k$ and $l$ are arbitrary separation constants.
These are, however, fixed by the boundary conditions:
Since $g_i(r) \, {\tilde g}_i(\theta) \to 1$ as $r \to \infty$,
each function must approach a constant, which we can require to be one, 
as $r \to \infty$:  i.e.\
$g_i(r)  \to 1$ and ${\tilde g}_i(\theta) \to 1$. 
The latter requirement dictates that ${\tilde g}_i(\theta) = 1$, 
so that the angular part is trivial.
This fixes the separation constants completely: $k=4$ and $l=5$.
(We note that this is also self-consistently required by the radial parts of 
\sepA\ and \sepB.)

Thus, we are left with the following coupled, linear, first order,
ordinary differential equations for $g_1(r)$ and $g_2(r)$:
\eqn\odeA{
g_1(r) = f(r) \, g_2(r) + {1 \over 4} r \, 
{d \over dr} \(  f(r) \, g_2(r) \) }
\eqn\odeB{
g_2(r) =  g_1(r) + {1 \over 5} r \, 
{d \over dr}  g_1(r) }
with the asymptotic boundary conditions $g_i(r) \to 1$ as $r \to \infty$.
Ordinarily, one would expect to be able to specify both $g_1$ and
$g_2$ at $r=r_+$ and then integrate out to infinity. One could then hope
to choose these two initial conditions to satisfy the two boundary conditions.
However, \odeA\ implies the following constraint at the horizon
(using the fact that $f(r_+) = 0$):
\eqn\constr{ g_1(r_+) =  {7 \over 4}  \, g_2(r_+)} 
so the solutions are determined by only one free parameter. Nevertheless,
it is still possible to satisfy both boundary conditions. This is most 
easily seen by substituting 
\odeB\ into \odeA\  to obtain a decoupled,
second order equation for $g_1(r)$:
\eqn\sode{
f(r) \, g_1''(r) + \( {10 \over r} \, f(r) + f'(r) \) g_1'(r) + 
\( {20 \over r^2} \, (f(r)-1) + {5 \over r} f'(r) \) g_1(r) 
= 0
}
The asymptotic form of this equation is 
\eqn\asympeq{
g_{1}''(r) + {10 \over r} \, g_{1}'(r) = 0,}
so as $r \to \infty$, we have
$g_1(r) \sim $ const + $O(1/r^9)$. There is only a one parameter family of
solutions to the second order equation \sode\ which are regular at the horizon
since $f(r_+)=0$ implies
\eqn\bc{
{g_1}'(r_+) = - {15 \over 7 r_+} \, g_1(r_+) }
So given $g_1(r_+)$, we get a unique solution of the second order equation
\sode. We can clearly rescale $g_1(r_+)$ so that $g_1\rightarrow 1$
at infinity. The function $g_2$ is then completely determined by \odeB,
but fortunately it automatically
satisfies the right boundary condition, $g_2\rightarrow 1$ asymptotically.
This shows that a solution satisfying all
boundary conditions does exist.

\ifig\soln{Solutions  $g_1(r)$ (dotted line) and 
$g_2(r)$  (dashed line), and their asymptotic value 
of one  (solid line)
for a 10-D \BH\ with radius $r_+ = 1$}
{\epsfxsize=9.5cm \epsfysize=5.5cm \epsfbox{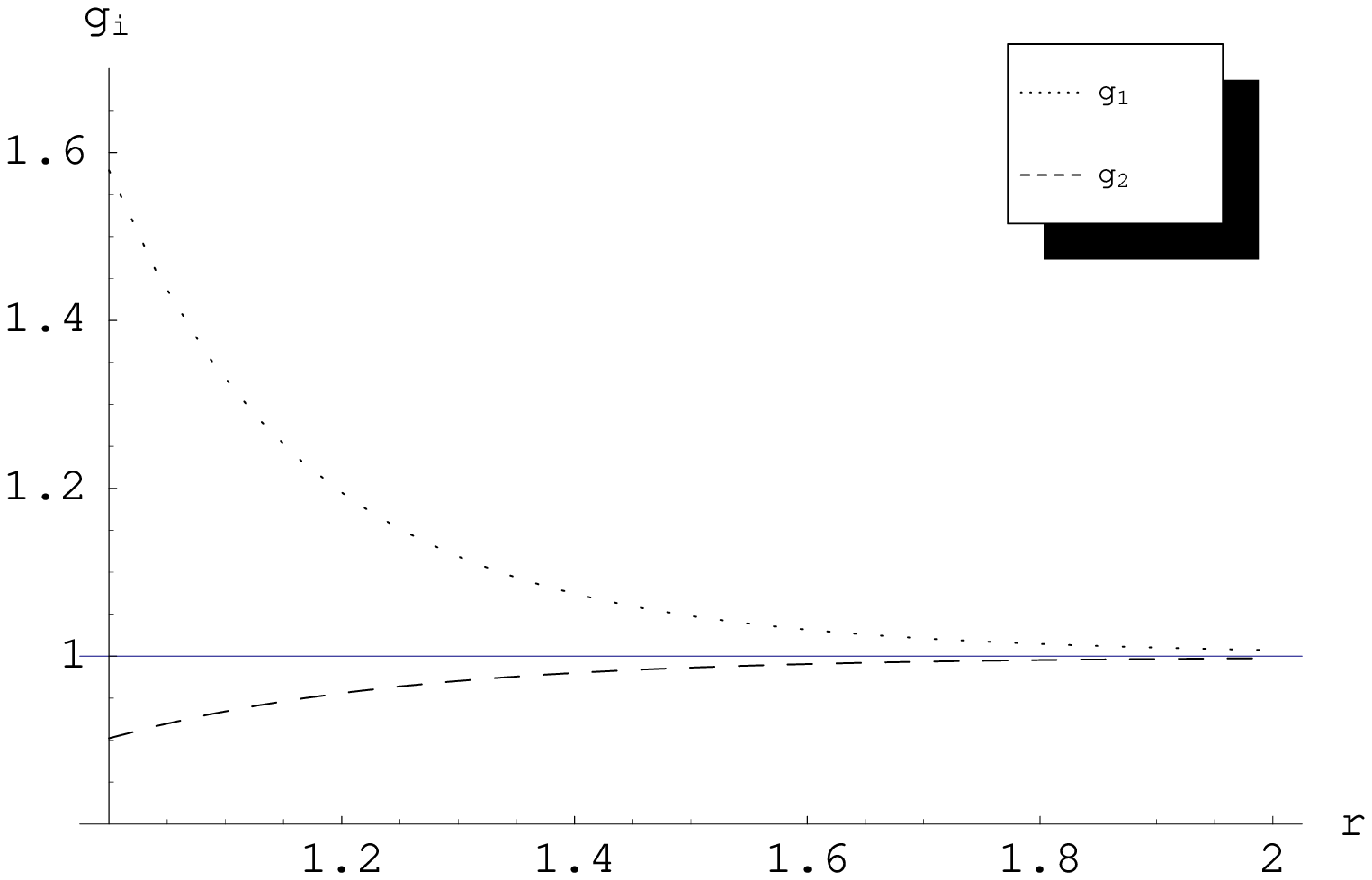}}
Although we have not found the solution analytically, one can
easily find it numerically.\foot{
However, we cannot integrate the solution directly from the horizon,
since  $g''_1(r_+)$ is undetermined there due to the factor of
$f(r) =0 $ at $r=r_+$.  Instead, we must obtain new ``initial conditions''
near the horizon, at $r=r_+ + \varepsilon$.
}
A plot of the solution is shown in
\soln.  We see that $g_1$ is enhanced and $g_2$ is slightly suppressed at the
horizon, while both functions 
asymptote to the correct value, $g_i \to 1$.

Even though we have found static solutions for a test five-form field
strength in a 10-D Schwarzschild background (given by \Fgdr), 
$F$ does not vanish at the horizon. So to ensure that the solution
remains static when the backreaction is included, we need to check that
there is no energy flux crossing the horizon.
By the Raychaudhuri equation \wald, 
the horizon area can remain constant only
if $R_{ab} k^a k^b = 0$, where $k^a$ denotes the null generators
of the horizon, $k^a = \( {\partial \over \partial t} \)^{a}$.
Thus, a static configuration must satisfy 
$T_{ab} k^a k^b = 0$ at the horizon.
One can easily show that this is indeed the case for our solution:
\eqn\tab{ T_{ab} k^a k^b 
 \propto k^a  F_{a c d e m} k^b  F_{b}^{\ c d e m} }
and from \Fgdr, we have (in component notation)
\eqn\kfacdm{k^a  F_{a  c d e m} \propto 
g_1(r) \, r^{3} \, \sin^{4} \theta \, 
(dr)_{[c} (d\Omega_{3})_{d e m]} +
g_2(r) \, r^{4} \, f(r) \, \sin^{3} \theta \, \cos \theta \,
(d\theta)_{[c}  (d\Omega_{3})_{d e m]}}
Contracting over $c,d,e,$ and $m$ yields
\eqn\squaref{ k^a  F_{a c d e m} k^b  F_{b}^{\ c d e m}  \propto 
f(r) \, g_1^2(r) \, \sin^{2} \theta +
f^2(r) \, g_2^2(r) \, \cos^{2} \theta }
which clearly vanishes at the horizon, since $f(r_+)=0$
and $g_i(r_+)$ remain finite.
Hence 
\eqn\hor{
T_{ab} k^a k^b = 0}
 is indeed satisfied at the horizon.

%

So far, we have considered a five-form field strength in the
presence of a small 10-D \BH\ in asymptotically
\AfSf\ spacetime.
We now check that the arguments of the preceeding section
also apply to  the other cases of interest for the AdS/CFT correspondence.

We start with the 11-D supergravity solutions \AfSs\ and \AsSf.
For conciseness, we combine these into the general case of
$AdS_p \times S^q$, where $(p,q) = (4,7)$ and $(7,4)$.
Here the logic of the argument is slightly different
from the previous case, since dimensionally the field strength cannot be
self-dual.
Nonetheless, we shall see that the final differential equations 
are very similar to \odeA\ and \odeB\ (and are in fact identical
if we set $p=q=5$).  
This will allow us to apply the same arguments as above
to prove the existence of a static solution satisfying the correct 
boundary conditions.

As in the preceeding discussion,
we start with the metric in global AdS coordinates:
\eqn\AdSpSq{
ds^2  \, = -\({\rho^2 \over R^2} + 1\) \, dt^2 + 
{d\rho^2 \over {\rho^2 \over R^2} + 1}
+ \rho^2 \, d\Omega_{p-2}^2
+ d\chi^2 + (\alpha R)^2 \, 
\sin^2 {\chi \over \alpha R}  \, d\Omega_{q-1}^2
}
where $\alpha$ is a numerical constant, corresponding to the
ratio of the size of the sphere to the size of AdS for the given
supergravity solution 
($\alpha={1 \over 2}$ for \AsSf, and $\alpha =2$ for \AfSs).
The flat space approximation ($\rho, \chi \ll R$) of the metric
and the corresponding volume form are given by 
\eqn\Mpq{
ds^2  \, = - dt^2 + d\rho^2
+ \rho^2 \, d\Omega_{p-2}^2
+ d\chi^2 + \chi^2 \,  d\Omega_{q-1}^2
} 
\eqn\epspqr{
\eps_{p+q} =  \rho^{p-2} \, \chi^{q-1} \, 
 dt \wedge d\rho \wedge  d\Omega_{p-2} \wedge  d\chi \wedge  d\Omega_{q-1}
}
Hence, the $p$-form field strength and its $q$-form dual 
in this region are simply
\eqn\Fp{
F_{(p)} = - \rho^{p-2} \, dt \wedge d\rho \wedge  d\Omega_{p-2}
}
\eqn\Fq{
\ast F_{(q)} = \chi^{q-1}  \, d\chi \wedge  d\Omega_{q-1}
}

Now, we use the change of coordinates \rhochi\ to  
write the $(p+q)$-dimensional Schwarzschild metric in the form 
\eqn\schwpq{
ds^2  \, = - f(r)  \,  dt^2 + f^{-1}(r)  \,  dr^2
+ r^2 \, \( d\theta^2 + \sin^2 \theta \, d\Omega_{p-2}^2
+ \cos^2 \theta \,  d\Omega_{q-1}^2 \)
}
where $f(r) \equiv 1 - \( {r_+ \over r} \)^{p+q-3}$.
Since the full $(p+q)$-dimensional volume form is independent of $f$,
it can be obtained from
\rhochi\ and \epspqr
\eqn\epspqrr{
\eps_{p+q} = 
(-1)^{p-1} \, r^{p+q-2} \, \sin^{p-2}\theta \, \cos^{q-1}\theta \, 
 dt \wedge dr \wedge d\theta  \wedge d\Omega_{p-2} \wedge  d\Omega_{q-1}
}

Up till now, everything was just a simple generalization of the
\AfSf\ case. 
However, the general $p$-form in the presence of the localized \BH, 
which is consistent with all the symmetries now has only two
arbitrary functions,
$$F_{(p)} = -g_1(r,\theta) \, r^{p-2} \, \sin^{p-1} \theta \,
 dt \wedge dr \wedge  d\Omega_{p-2}$$
\vskip -0.3 in
\eqn\Fpg{- g_2(r,\theta) \, f(r)  \, r^{p-1} \, 
\sin^{p-2} \theta \, \cos \theta \,
 dt \wedge d\theta \wedge  d\Omega_{p-2}}
$g_1(r,\theta)$ and  $g_2(r,\theta)$ are smooth everywhere and chosen such that
they satisfy the simple flat space boundary condition\foot{
The function $f(r)$ was inserted into the second term for later convenience.
(Note that $f(r) \to 1$ as $r \to \infty$, so the asymptotic boundary conditions
remain uneffected.)}
$g_1(r,\theta) \to 1$ and  $g_2(r,\theta) \to 1$ as $r \to \infty$.
(The other two terms which appeared in  \Fgr\ for $F_{(5)}$ 
are not consistent with the dimensionality:
they are $q$-forms rather than $p$-forms.)
The dual $q$-form is then
\eqn\Fqg{
\ast F_{(q)} = -g_1(r,\theta) \, r^{q} \, \sin \theta \, \cos^{q-1} \theta \,
 d\theta \wedge  d\Omega_{q-1}
+ g_2(r,\theta)  \, r^{q-1} \,  \cos^{q} \theta \, dr \wedge  d\Omega_{q-1}}

The differential equations for $g_1$ and  $g_2$ 
are obtained by the  condition
that both the $p$-form field strength and its dual $q$-form must be
closed, i.e.\ $d F_{(p)}=0$ and $d\! \ast\!\! F_{(q)}=0$.
Canceling out the angular dependence 
(using the same separation of variables procedure as before) yields
\eqn\odep{
g_1(r) = f(r) \, g_2(r) + {1 \over p-1} r \, \partial_r \(  f(r) \, g_2(r) \) }
\eqn\odeq{
g_2(r) =  g_1(r) + {1 \over q} r \, \partial_r  g_1(r) }
Note that, as advertised, for $p=q=5$, 
these equations are  identical to \odeA\ and \odeB\ obtained above.
The second order ODE for $g_1(r)$ is
\eqn\sodepq{
f(r) \, g_1''(r) + \( {p+q \over r} \, f(r) + f'(r) \) g_1'(r) + 
\( {q(p-1) \over r^2} \, (f(r)-1) + {q \over r} f'(r) \) g_1(r) 
= 0
}
The asymptotic form of this equation,
\eqn\aspq{
g_{1}''(r) + {p+q \over r} \, g_{1}'(r) = 0}
implies $g_1(r) \sim$ const + $O(1/r^{p+q-1})$ 
as $r \to \infty$.

Solutions which are regular at the horizon are again determined by one 
parameter since
\eqn\bc{
{g_1}'(r_+) = -{q \over r_+} \, \( {q-2 \over p+q-3}\) \, g_1(r_+) }
One can choose this parameter so that $g_1 \to 1$ asymptotically. 
Then $g_2$ is uniquely determined by \odeq\ and automatically satisfies
the right boundary condition $g_2\to 1$.

One can similarly show that the same conclusion will hold for
another case of interest for the AdS/CFT duality, 
namely in IIB supergravity on $AdS_3 \times S^3 \times T^4$.
In this case, the 4-torus decouples, and by a similar
procedure as for \AfSf, we arrive at equations \odep\ and \odeq,
with $p=q=3$.

In all of the above cases one can easily verify that there is no
flux of energy across the event horizon,  $T_{ab} k^a k^b = 0$. 
(Note that although
the stress tensor now has a nonzero trace, the term proportional to the metric
does not contribute when contracted
with the null vectors $k^a k^b$.)
So the solutions will remain static when backreaction is included.
Thus, we have shown that in all the relevant cases, 
a small \BH\ in $AdS_p\times S^q$ 
can indeed be approximated by a $(p+q)$-dimensional Schwarzschild solution.
The Ramond-Ramond fields will be distorted, but will remain static and
not cause the black hole to grow. 
In retrospect, it is perhaps not surprising that static solutions
do exist, since they can be viewed as higher dimensional generalizations
of a \BH\ in a background magnetic field \ernst.

It should be noted that the validity of the Schwarzschild approximation 
does not imply that all small black holes
will Hawking evaporate. As noted in \gary, 
if one fixes the total energy, the asymptotic AdS boundary conditions ensure
that certain small black holes can be in stable
equilibrium with their own Hawking radiation.
However, sufficiently small black holes will still evaporate, so the AdS/CFT
correspondence leads one to believe that this evaporation can be described
by a unitary evolution.

\vskip 1cm

\centerline{\bf Acknowledgements}

This work was supported in part by NSF Grant PHY95-07065.
%
%

\listrefs
\end